\documentclass[onecollarge,natbib]{svjour2}
\bibpunct{[}{]}{;}{n}{}{,} % to get "[numbered]" references from natbib
\smartqed  % flush right qed marks, e.g. at end of proof
\usepackage{graphicx}
\journalname{Few-Body Systems}
\begin{document}

\title{Tunneling of atoms, nuclei and molecules%\thanks{Grants or other notes
}


\author{C.A. Bertulani
}


\institute{C.A. Bertulani \at
              Department of Physics and Astronomy, Texas A\&M University-Commerce, Commerce TX 75429, USA  \\
              Tel.: +1-903-886-5882\\
              Fax: +1-903-886-5480 \\
              \email{carlos.bertulani@tamuc.edu}           %  \\
}

\date{Received: date / Accepted: date}

\maketitle

\begin{abstract}
This is a brief review of few relevant  topics on tunneling of composite particles and how the coupling to intrinsic and external degrees of freedom affects tunneling probabilities. I discuss the phenomena of resonant tunneling, different barriers seen by subsystems, damping of resonant tunneling by level bunching and continuum effects due to particle dissociation.
 
\end{abstract}

\section{Particles moving in mysterious ways}
\label{intro}

\paragraph{History - }The tunnel effect is one of the most subtle phenomenon explained by  quantum mechanics, responsible e.g., for the existence of stars and ultimately for the existence of life \cite{Hoy54}. The first application of particle incursion into classically forbidden regions was done in nuclear physics. Nuclei, such as $^{210}$Po, emit $\alpha$-particles by tunneling through the Coulomb barrier. This process lacked physical explanation until George Gamow used the tunneling theory to calculate $\alpha$-emission half-lives \cite{Gam28,GC28}. Tunneling is now a well known physical process that has been incorporated in our everyday lives due to the increasing miniaturization in electronics such as microchips. There have been important Nobel Prizes related to the straightforward use of this effect in industry, such as the tunnel diode \cite{Esa73} or the scanning tunneling microscope \cite{BR86}.

\paragraph{Resonant tunneling - }Resonant tunneling is a particular kind of tunneling effect, frequently applied  to miniaturization such as the resonant diode tunneling device. In its simplest form, resonant tunneling occurs when a quantum level in one side of a barrier has an energy match with a level on the other side of it.  If this occurs in a dynamical situation, tunneling is enhanced. In certain situations, the transmission probability is equal to one and the barrier is completely transparent for particle transmission.  In the resonant diode tunneling device, two semiconductor layers sandwich another creating a double-humped barrier which enables the existence of  quantum levels within.  On both sides of the barrier, electrons fill a conducting band on two outside semiconductors. A potential difference between these outer layers allows one conducting band to rise in energy. Resonant tunneling leaking occurs to the levels inside the double-hump barrier \cite{Esa73}. The net effect is the appearance of a current which can be fine tuned. Resonant tunneling devices are compact and allow a fast response because the tunneling between the thin double-humped barrier is a very fast process.

\begin{figure*}
\centering
  \includegraphics[width=0.45\textwidth]{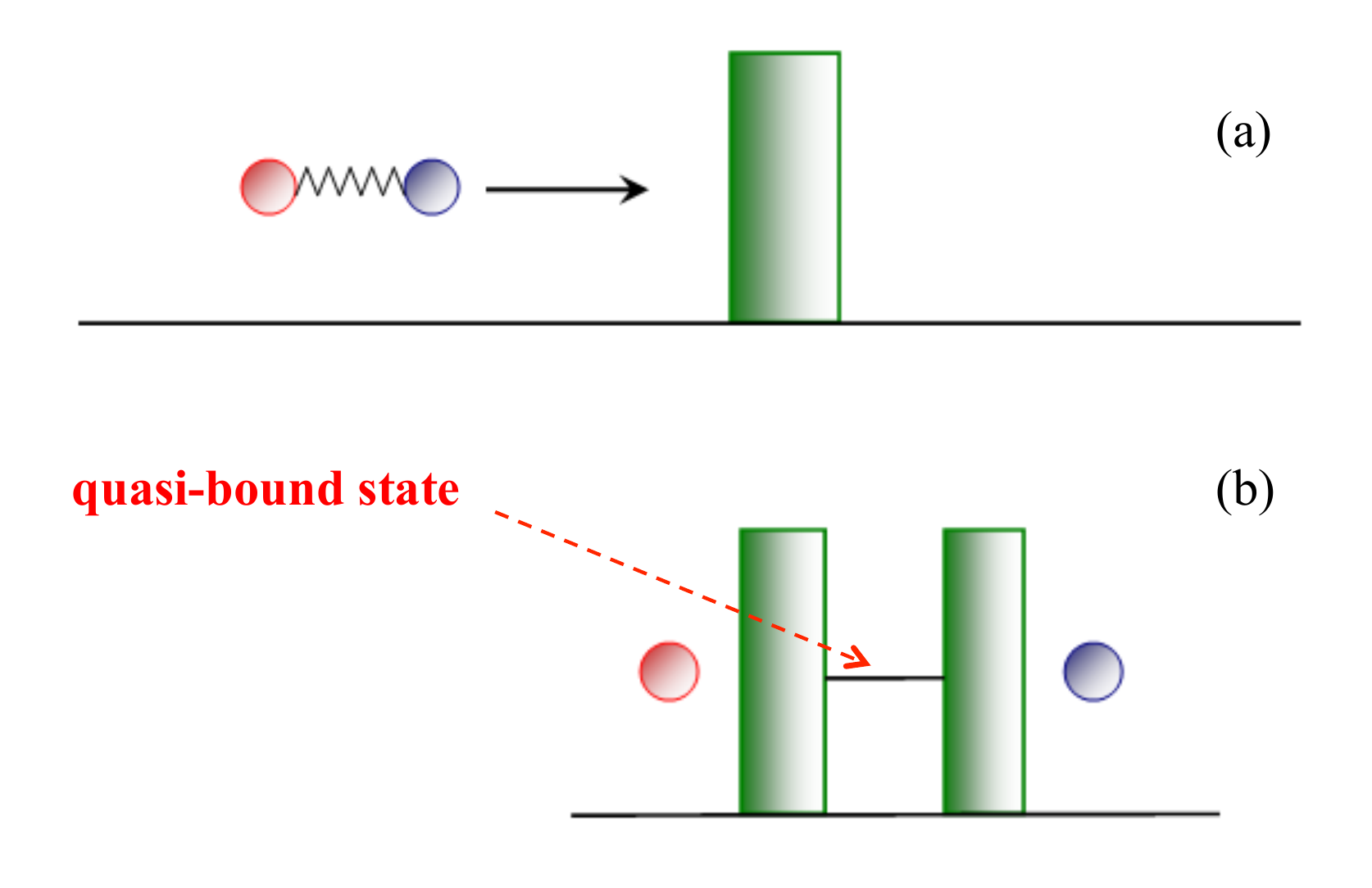}
\ \ \ \  \ \ \ \ \  \ \ \ \ \ \ \ \includegraphics[width=0.39\textwidth]{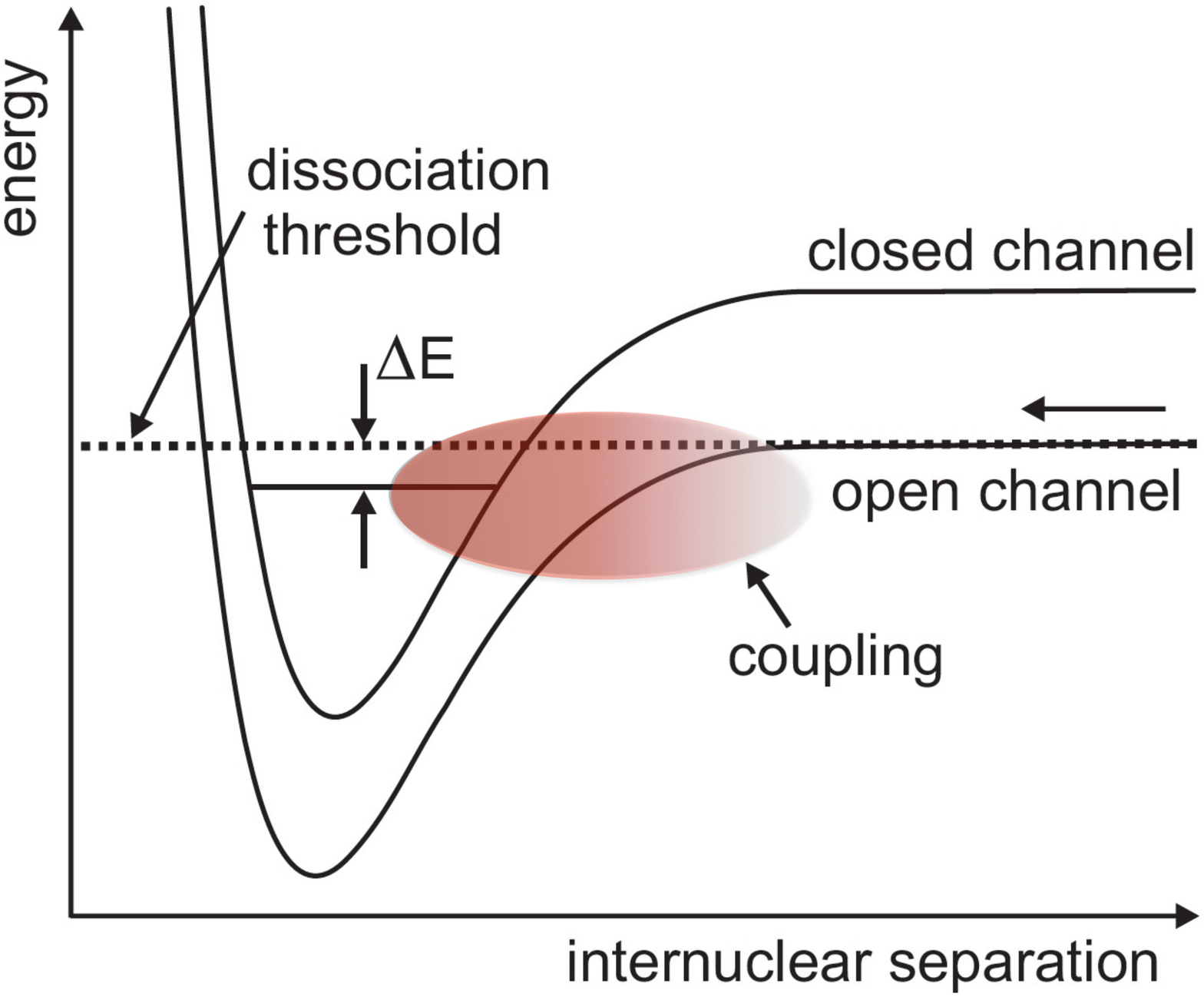}
\caption{{\it Left:} Tunneling of a composite particle through a barrier is often equivalent to the tunneling of a single particle through a double-humped barrier. {\it Right:} A Feshbach resonance appears when a preferred tunneling is induced by the coupling of a reaction channel to one or more intrinsic degrees of freedom.}
\label{comptun}       % Give a unique label
\end{figure*}

\paragraph{Composite particles - }Resonant tunneling is not constrained to a particle tunneling though a double-humped barrier. An equivalent process occurs when a composite particle tunnels through a single barrier, if each subsystem composing the particle is allowed to tunnel independently or when the interaction reveals a priori unknown energy states. Then the process is equivalent to the tunneling of one single particle though a double-humped barrier, and resonant tunneling proceeds through the appearance of pseudo, quasi-bound, states (see Figure \ref{comptun}, left). This effect occurs frequently in atomic, molecular and nuclear systems.  A good example is the  fusion of loosely-bound nuclei \cite{Bert11}. It also relates to Feshbach resonances when a system uncovers a state in a (closed) channel which is not the same (open) channel where it sits in. This happens because the coupling with at least one internal degree of freedom helps the reaction to proceed via resonant tunneling (see Figure \ref{comptun}, right). 

\begin{figure*}
\centering
  \includegraphics[width=1.05\textwidth]{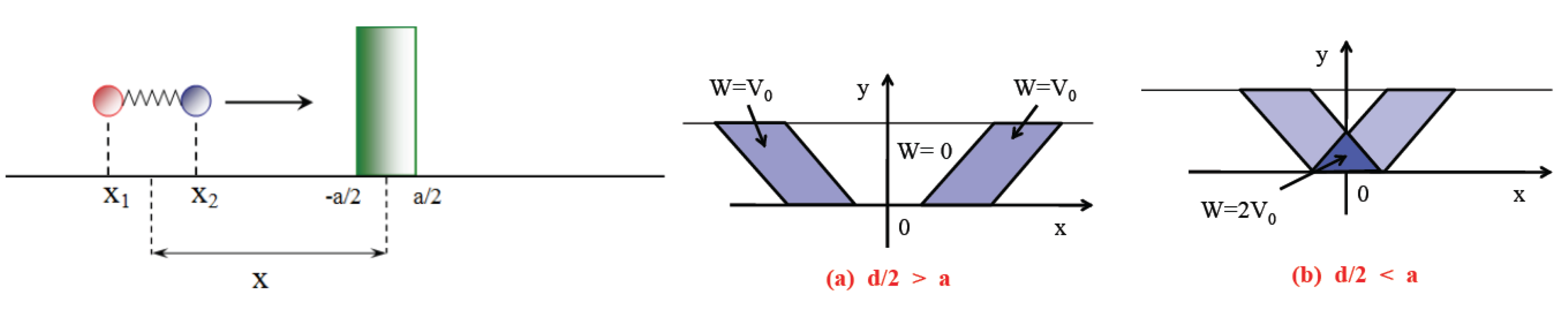}
\caption{{\it Left:} Tunneling of a composite particle through a single barrier. {\it Right, (a) and (b):} The effective potential for the center of mass motion of the system.}
\label{dh}       % Give a unique label
\end{figure*}

In the next Section I give several examples of how intrinsic degrees of freedom manifest  in subtle ways, leading to tunneling enhancement or suppression.  As the subject is very vast and general, I will concentrate on examples that I had the opportunity to work with during the last decades. They range from particle to molecular physics. Some common characteristics are easy to understand, others not so much. An example of the last case if given for diffusion and dissociation of molecules in optical lattices. 

\section{Tunneling of composite particles}

Most theoretical problems involving tunneling are not amenable to analytical solutions. There are several computational methods available for them. Below I will just mention a very well-known method, used to generate some of the calculations displayed in the figures of this review.

\paragraph{Stalking microscopic particles - }It is interesting to study the time-dependence of tunneling, although it is prohibitive for most cases of interest due to the long tunneling times, e.g., in nuclear $\alpha$-decay processes. The time evolution of a wave function on a space lattice can obtained by solving the Schr\"odinger equation by a finite difference method. The wave function $\Psi(t+\Delta t)$ can be calculated from the wave function at time $t$, $\Psi(t)$, by applying the unitary time evolution operator, $U$, i.e., $\Psi(t+\Delta t) = U(\Delta t, t) \Psi(t)= \exp{\left[ -iH\Delta t/\hbar\right]} \Psi(t)$, where $H$ is the system Hamiltonian.  For a small time step $\Delta t$,  one can use the unitary operator approximation, valid to order $(\Delta t)^2$,
\begin{equation}
U (t + \Delta t) = {1+(\Delta t/2i\hbar)H(t) \over 1-(\Delta t/2i\hbar)H(t)} \label{Udt}.
\end{equation}
Other approximations for $\exp{\left[- iH\Delta t/\hbar\right]} $ such as the Numerov algorithm, etc., can also be used. Any of these numerical procedures requires carrying out matrix multiplications and inversions at each iteration and the number of operations grows very fast with the number of points in the coordinate lattice \cite{Var62,PTV07}. It also increases with the number of particles in the system at hand, becoming numerically demanding for large fermionic or bosonic systems \cite{Ste15}. But the most difficult problem is related to the existence of times scales which differ by a large amount. For example, during alpha decay the decay time is often much larger (e.g., $10^{30}$ ) than the intrinsic time dependence of the alpha particle state in the nucleus. This makes it impossible to treat the problem in a time-dependent fashion. But there are ongoing theoretical efforts to handle the time-dependent problem with widely different time scales \cite{TBR91,PL04,BF06}.

\begin{figure*}
\centering
  \includegraphics[width=0.35\textwidth]{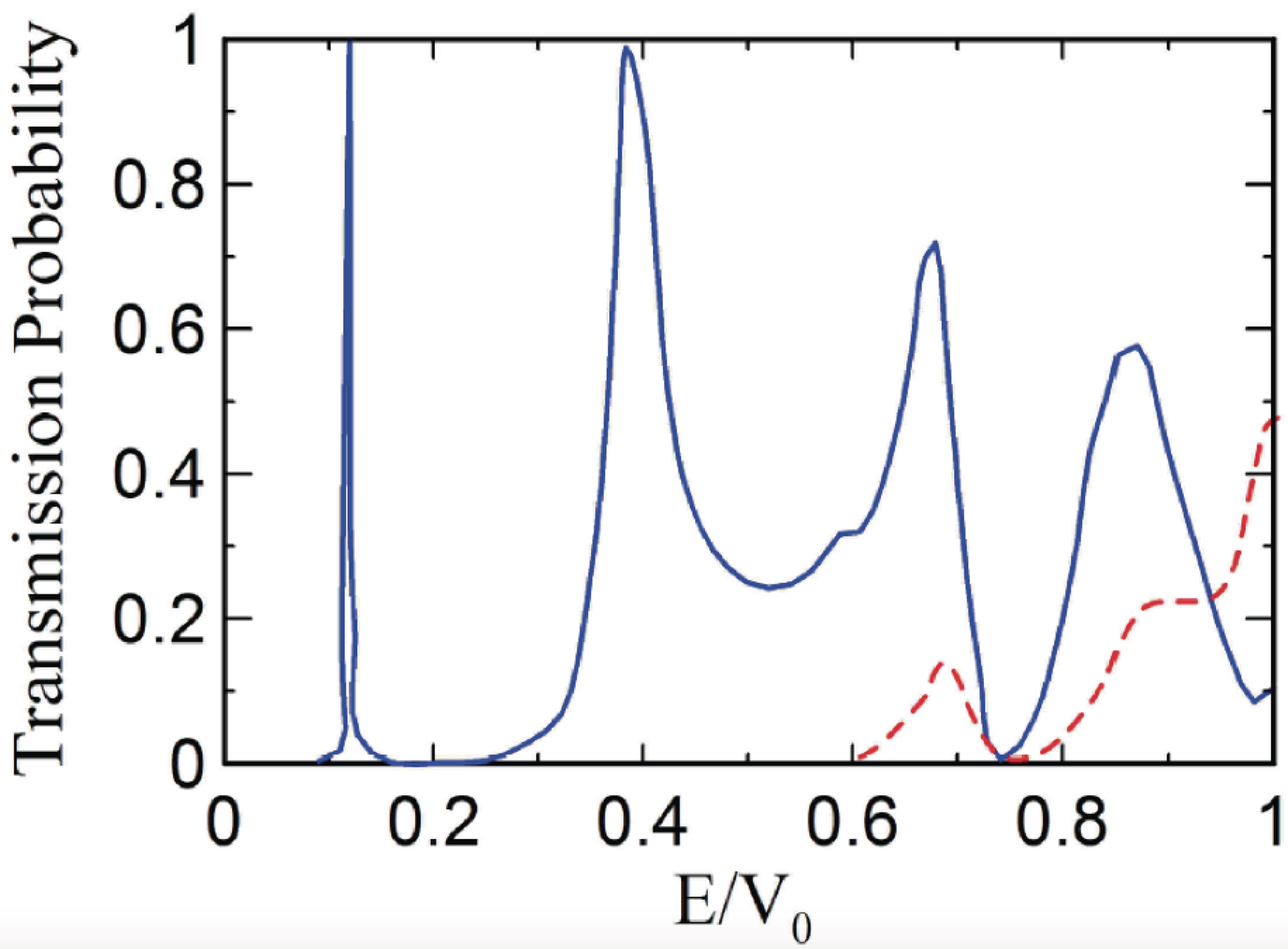}
  \ \ \ \  \ \ \ \ \  \ \ \ \ \ \ \ \includegraphics[width=0.34\textwidth]{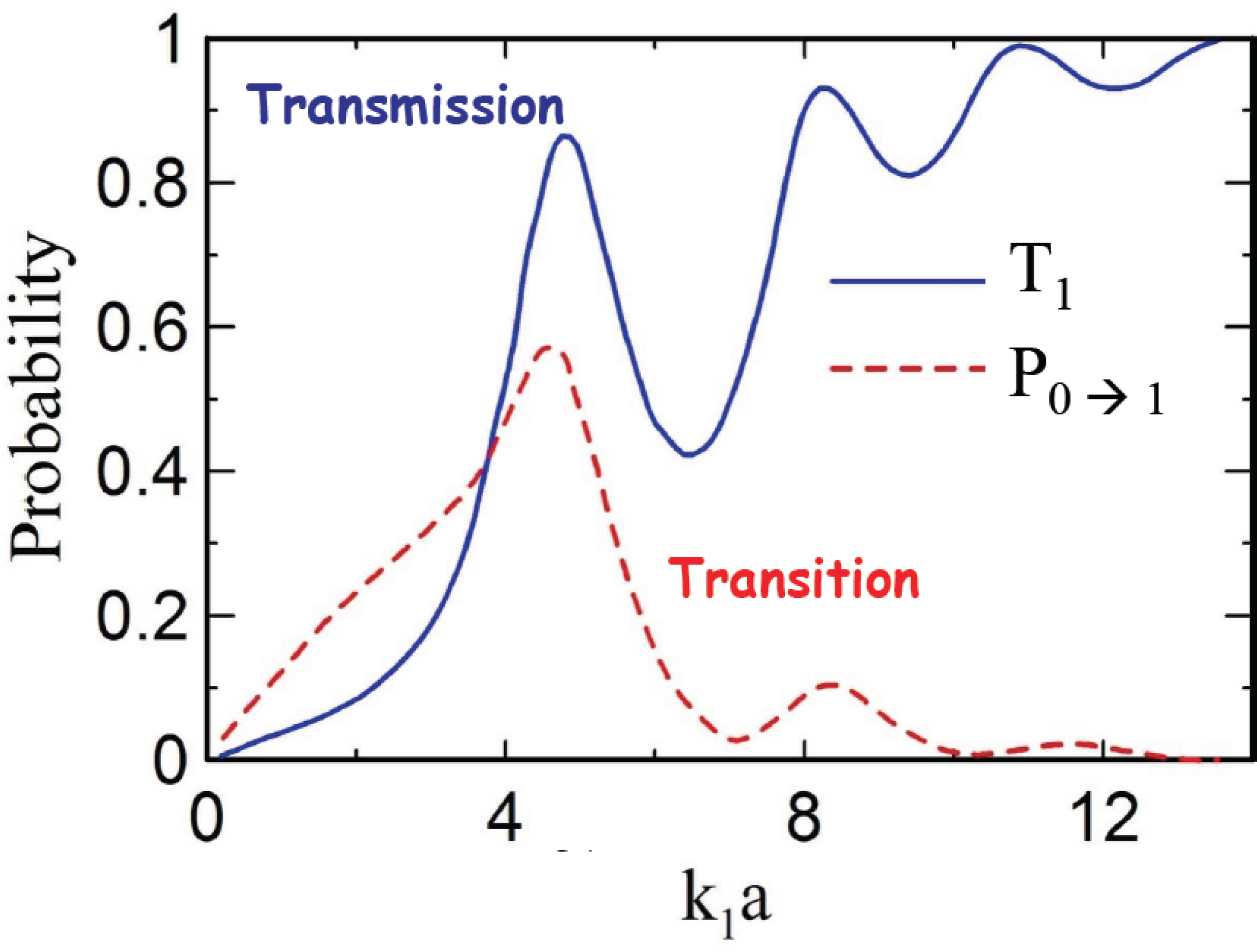}
\caption{{\it Left:} Transmission probabilities for a two-body system through a rectangular barrier \cite{SK94} as a function of the ratio of the center of mass energy and the barrier height $V_0$. The solid line is the transmission probability in which the system remains in the ground state and the dashed line is the for the case the system transits to an excited state. {\it Right:} Transmission probability of a molecule through a thin barrier as a function of the barrier width $a$ times the molecule momentum. The dashed curve is the transmission probability when the molecule is in an excited state after the transmission \cite{GS05}.}
\label{prob}       % Give a unique label
\end{figure*}

\begin{figure*}
\centering
  \includegraphics[width=0.31\textwidth]{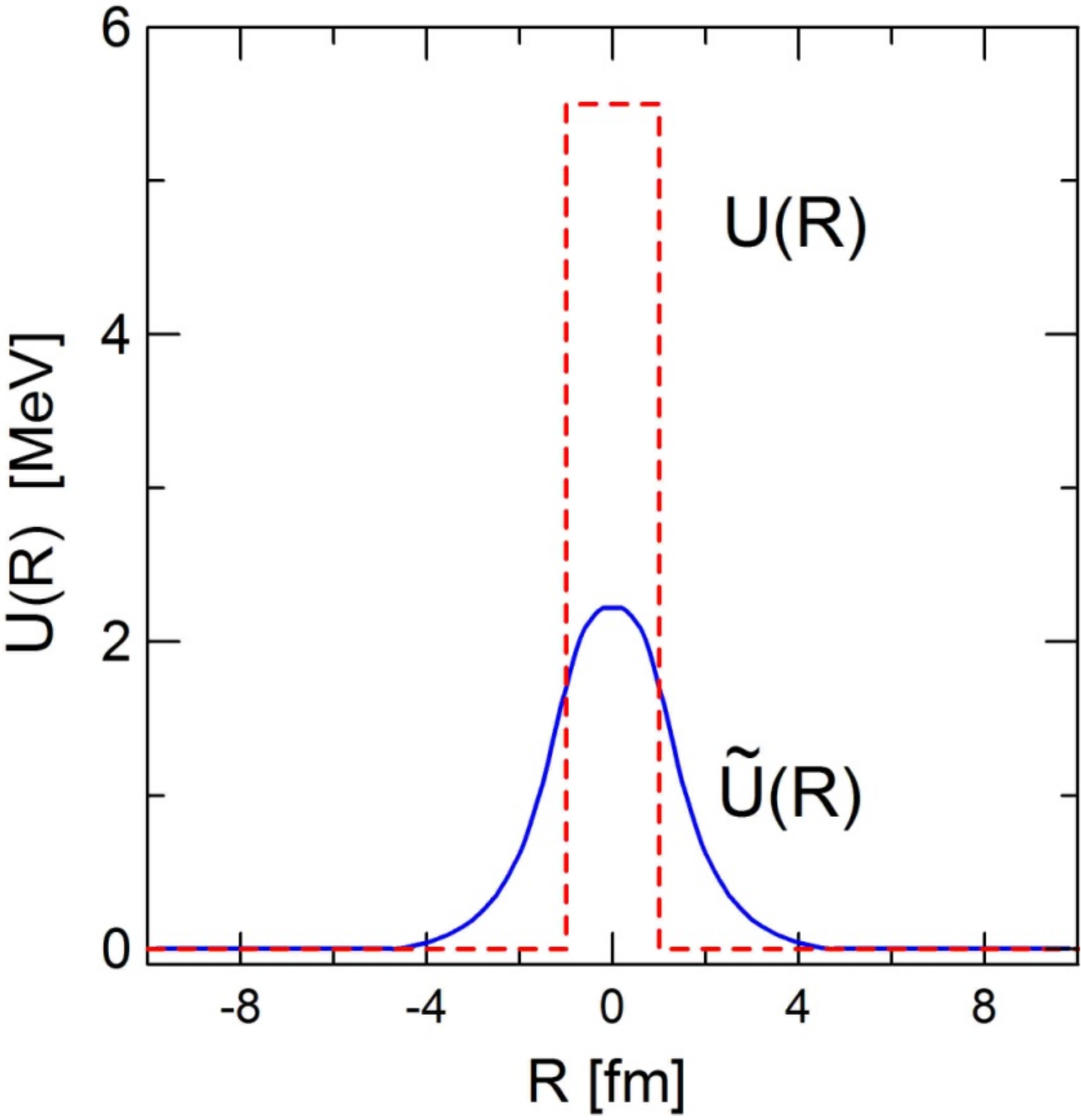}
  \ \ \ \  \ \ \ \ \  \ \ \ \ \ \ \ \includegraphics[width=0.31\textwidth]{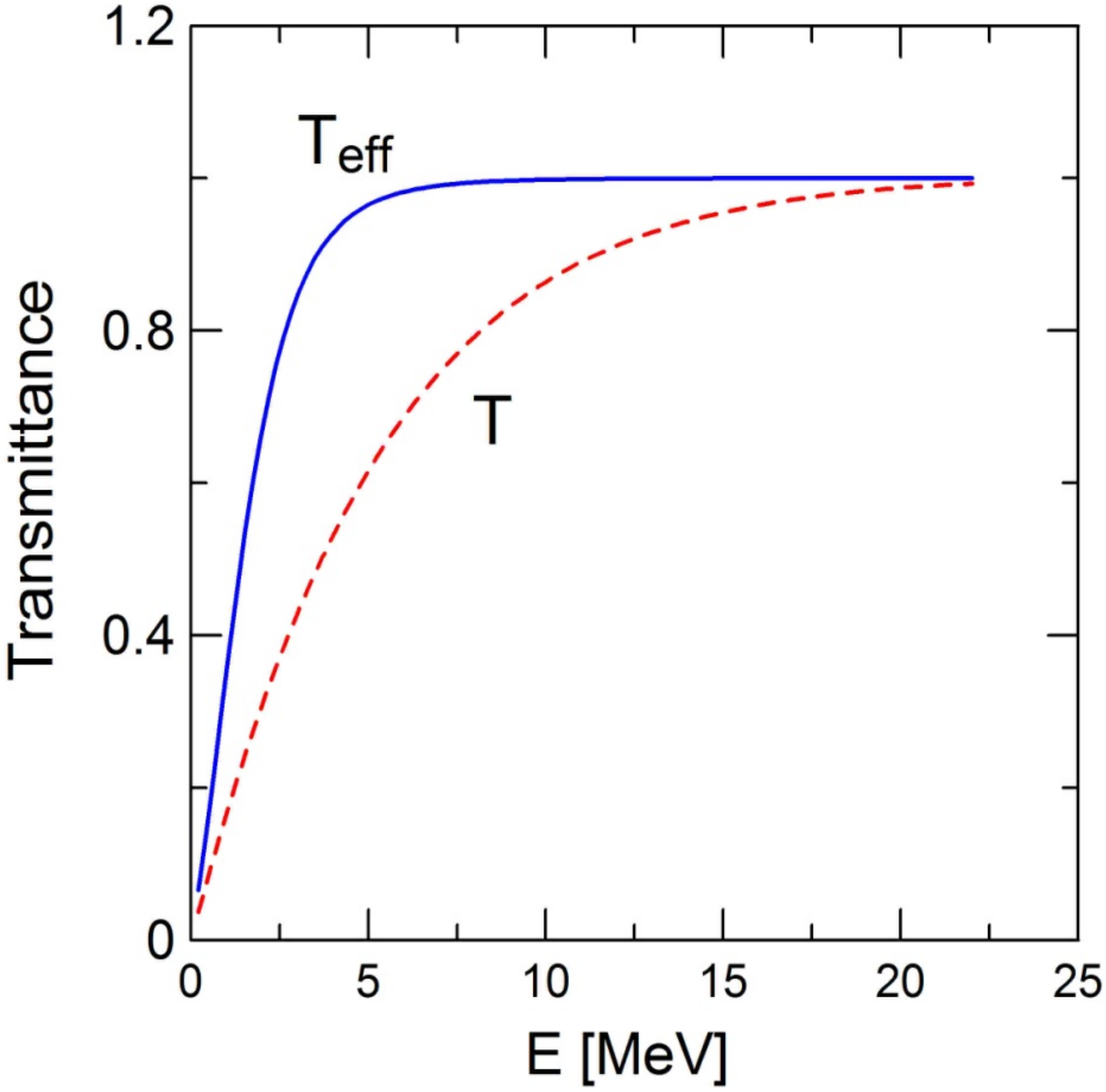}
\caption{{\it Left:} A schematic rectangular barrier simulates the proton-nucleus interaction for a deuteron-nucleus fusion reaction. The solid curve is the effective potential for the center of mass motion. {\it Right:} Transmission probability for the deuteron, with (solid) and without (dashed) consideration of coupling to intrinsic modes \cite{BFZ07}.}
\label{prob2}       % Give a unique label
\end{figure*}

As an application of the method described above, I will present below results for the tunneling of a two-body system through a simple rectangular barrier. 

\paragraph{Transmission of composite particles - } Assume a system of two bound particles with an attractive infinite square-well interaction of range $d$, being transmitted through a rectangular barrier of width $a$ and height $V_0$. By choosing the  center of mass $x=(x_1+x_2)/2$ and relative $y=x_1-x_2+d/2$ coordinates, it is straightforward to show that the center of mass wave equation has an effective double-humped potential $W(x,y)$. Figure \ref{dh} shows that for various values of the relative distance $y$, the effective potential $W$ is in fact a double-humped barrier \cite{Esa73,SK94}. This barrier can hold energy states that yields resonant tunneling if they are equal to the center of mass energy, as shown in Figure \ref{prob} for the dependence of the ratio of the bombarding energy and the barrier height \cite{Esa73,SK94}. 

The simple example mentioned above can be readily applied to a very practical case; that of a bound molecule transmitting through a barrier. This is illustrated below. 

\paragraph{Transmission of molecules - } Another way to enhance tunneling is allowing the system to make a transition from an excited state to a lower energy state. This has been shown analytically in Ref. \cite{BFZ07} for a two-spin system tunneling though a barrier. The transmission probability as a function of the barrier width also shows signs of resonant behavior, as observed in calculations for transmission of molecules though barriers \cite{GS05}. Figure \ref{prob} shows the transmission probabilities for a two-body system through a rectangular barrier \cite{SK94} as a function of the ratio of the center of mass energy and the barrier height $V_0$. The solid line is the transmission probability when the system remains in the ground state and the dashed line is when  the system transits to an excited state. The right hand side of Figure \ref{prob} shows the transmission probability of a molecule through a thin barrier as a function of the barrier width $a$ times the molecule momentum. The dashed curve is the transmission probability when the molecule is in an excited state after the transmission \cite{GS05}. By moving to its ground state the excited molecule increases its chance of transmission. On the opposite trend, the opening of the higher channels decreases the probability of transmission. 

The two-body tunneling problem also encounters abundant applications in nuclear physics, involving cluster-like objects. The simplest one is the deuteron, composed of a proton and a neutron forming a relatively loose-bound system ($E_B=-2.224$ MeV). Other common systems include those containing alpha particles. Tunneling involves sudden accelerations leading to gamma-emission in alpha tunneling, as described below.  

\paragraph{Fusion of loosely bound nuclei - } Loosely-bound systems also display interesting tunneling properties because they can end up in separate parts by reflection or transmission through a barrier. If the separate parts see a different barrier then the effective barrier for the center of mass tunneling can also be very different than the original one. A good example is the fusion of the deuteron with another nucleus in which case the proton-nucleus potential has a Coulomb repulsion part whereas the neutron-nucleus potential does not. As shown in Ref. \cite{BFZ07} the net effect is an enhancement of the transmission probability through the barrier (see Figure \ref{prob2}). Fusion of exotic, neutron rich, nuclei due to particle transfer, or tunneling, is now a very active area of research \cite{CDLH95,TKS93,HPCD92,HPCD93,DV95,Lem09,AV10,SS11} (for a review, see Ref. \cite{CGDH06}). 

An  intriguing question is  how much time a particle takes to tunnel a barrier \cite{LM94}. A Bremsstrahlung measurement following $\alpha$-decay \cite{Kas97} apparently paved the way to determine tunneling times by inspecting the interference pattern in the radiation spectra.  A time-dependent description of the emission process with a pre-formed $\alpha$-particle has shown that a model with a free, uncorrelated, $\alpha$-particle within the nucleus is incompatible with the observations for this process \cite{DPZ99}. The spectrum of Ref. \cite{Kas97} was also explained in a more traditional theoretical method \cite{PB98,Boi07,Jen08}. However, the question related to tunneling times is still fascinating, in particular when it comes to the possibility of time travel and the existence of time warps and worm holes \cite{Haw92}.

\begin{figure*}
\centering
  \includegraphics[width=0.41\textwidth]{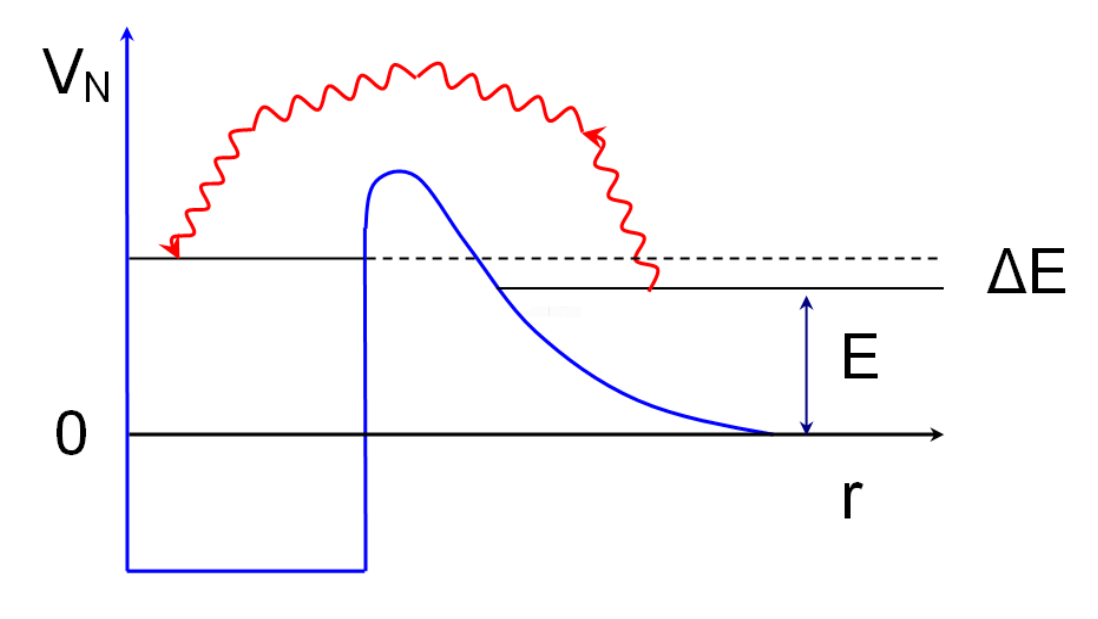}
  \ \ \ \  \ \ \ \ \  \ \ \ \ \ \ \ \includegraphics[width=0.44\textwidth]{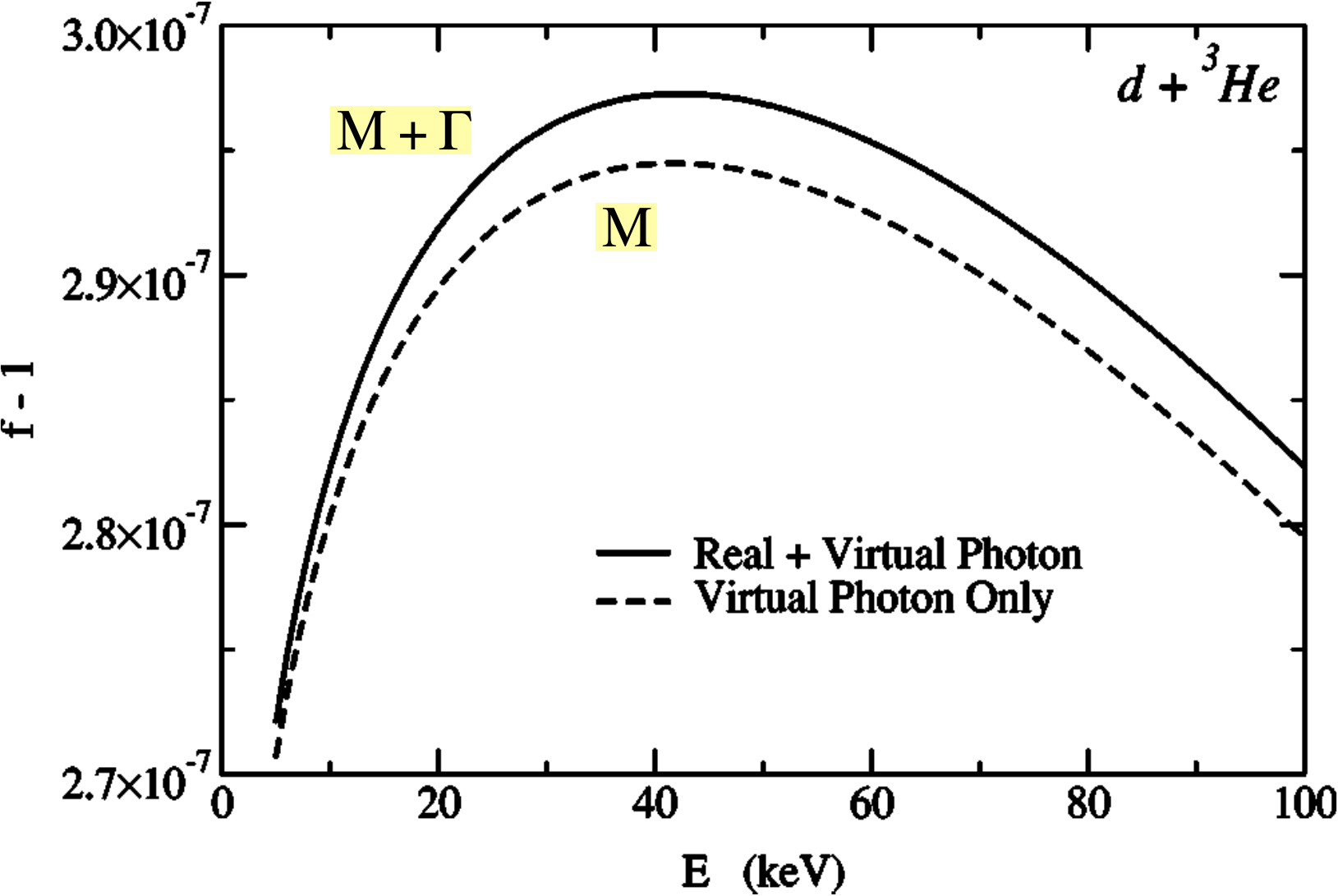}
\caption{{\it Left:} The quantum M\"unchhausen effect \cite{FZ99}. {\it Right:} Enhancement of fusion for the reaction d +  $^3$He due to couplings to real and virtual photons.}
\label{prob3}       % Give a unique label
\end{figure*}

The emission of radiation during tunneling is also manifest in virtual processes, a fascinating discussion of which is presented below.

\paragraph{The quantum M\"unchhausen effect - } Additional interactions during tunneling can lead to other subtle effects, e.g., as proposed in Ref. \cite{FZ99}. Since barrier transmission is proportional to an exponentially decreasing dependence with the particle energy, any small effect can change it dramatically. The so-called quantum M\"unchhausen\footnote{Baron M\"unchhausen is supposedly the greatest liar ever depicted in fiction literature and in movies. He once explained how he could save himself from sinking in quicksand by pulling his own hair up \cite{Ras70}.} effect assumes that a virtual photon emitted in one side of the barrier and caught on the other side can enhance barrier penetrabilities (Figure \ref{prob3}). This is allowable in quantum mechanics because a particle has an extended wave nature. The particle kinetic energy can momentarily increase (within bounds set by Heisenberg's uncertainty principle)  and tunneling can be enhanced. A non-relativistic reduction of the particle-radiation interaction yields \cite{FZ99}
\begin{equation}
H\Psi({\bf r}) + \int \Pi ({\bf r}, {\bf r}' ;E)\Psi({\bf r}') d{\bf r}'=E\Psi({\bf r}), \ \ \ \  {\rm where}  \ \ \ \ \Pi ({\bf r}, {\bf r}' ;E)= M ({\bf r}, {\bf r}' ;E)+i\Gamma ({\bf r}, {\bf r}' ;E).
\end{equation}
In this expression, $M$ includes the self-energy, mass renormalization and virtual photons, while $\Gamma$ includes the effects of  radiative decay width and  real photons. This effect has been studied further in Ref. \cite{HB02} and it was shown to be very small indeed for tunneling of stable nuclei. But an account of this effect for loosely-bound nuclei (and molecules) has not been explored at depth.

\begin{figure*}
\centering
  \includegraphics[width=0.35\textwidth]{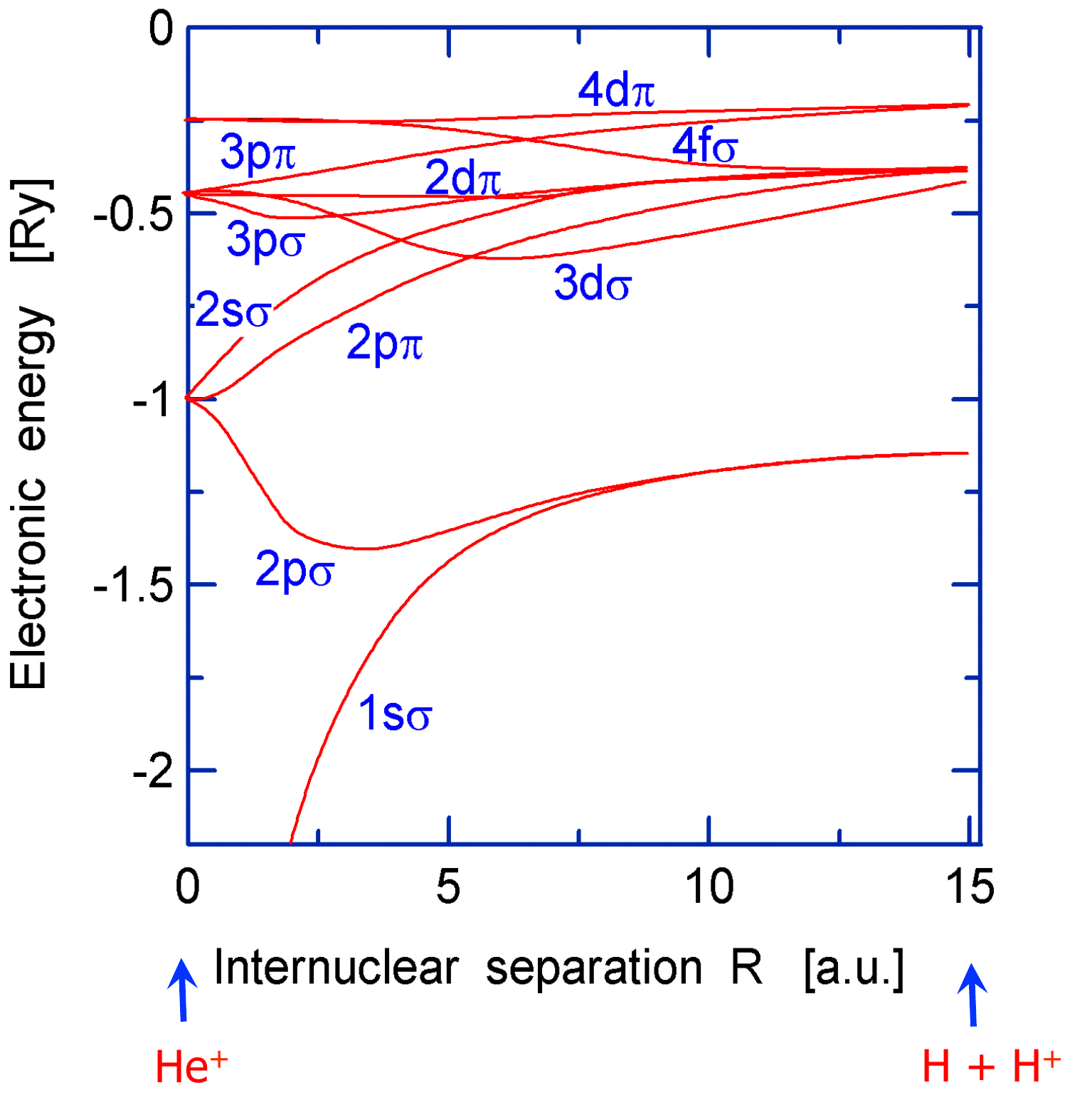}
  \ \ \ \  \ \ \ \ \  \ \ \ \ \ \ \ \includegraphics[width=0.31\textwidth]{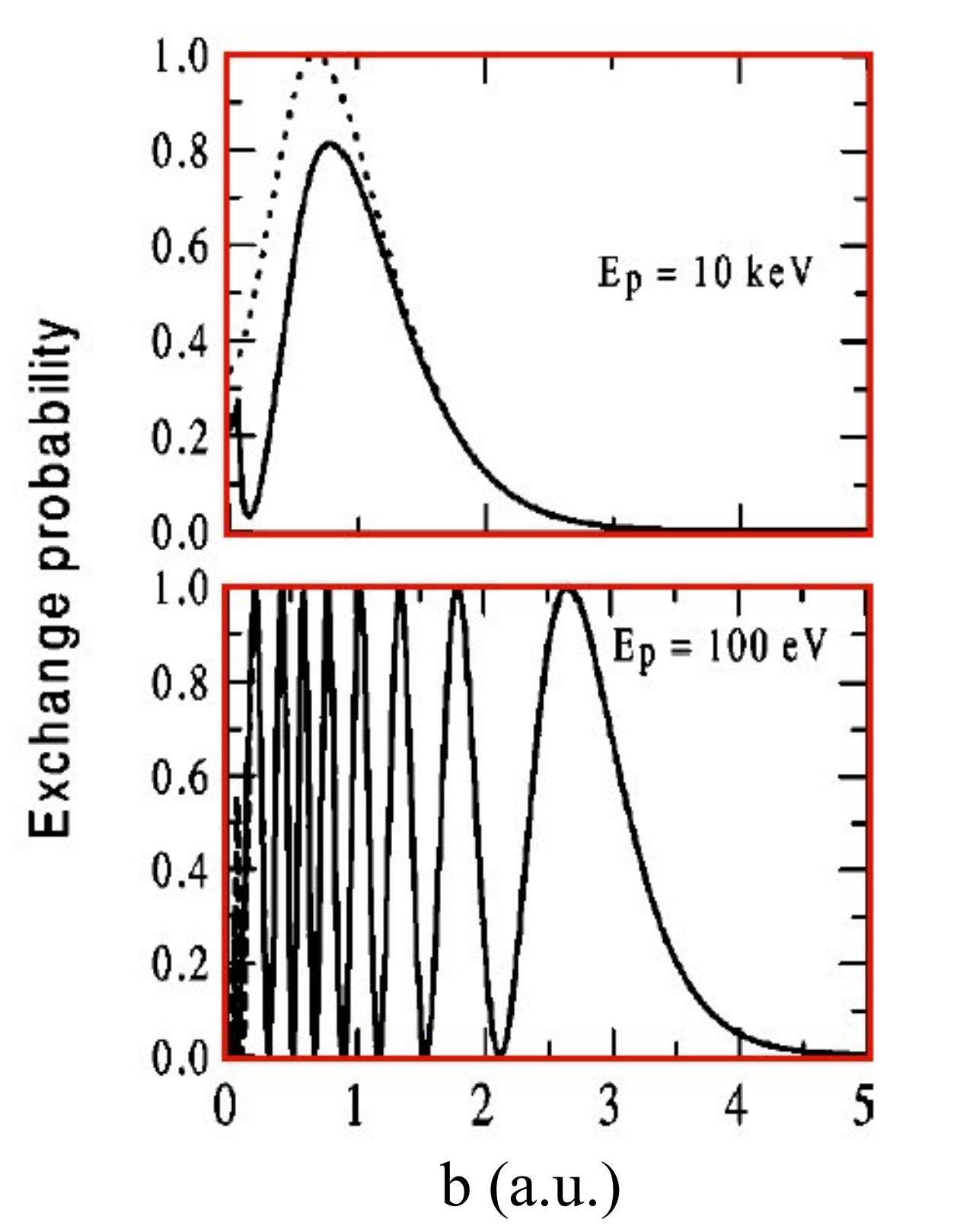}
\caption{{\it Left:} Energy levels of an electron as a function of the distance between a proton and a hydrogen atom. {\it Right:} The exchange probability as a function of the impact parameter for two proton kinetic energies. }
\label{phlev}       % Give a unique label
\end{figure*}

In atomic systems, electron tunneling is sometimes responsible for chemical bonding. But subtle effects are also present in dynamical situations, in which electron tunneling can be the cause of increased stopping of protons on a hydrogen target at extremely low energies, as discussed next.

\paragraph{Electron hoping induces stopping - }As discussed above, tunneling can be strongly affected by matching energy levels, the appearance of pseudo-levels during the tunneling process (Feshbach resonances), and by energy transfer to the relative motion by coupling to the internal degrees of freedom. Other effects appear when many energy levels are involved. Perhaps the best and clearer examples appear in atomic physics.  Stopping power, or the energy loss of a projectile per unit length, $S=- dE/dx$, is an important quantity for experiments at low bombarding energies. The energy at which a nuclear reaction occurs depends on the average energy loss of the projectile in an atomic target. Correcting for the stopping power is thus imperative for low energy experiments. At extremely low projectile energies, only a few phenomena are energetically viable such as the excitation of vibrations (plasmons) in the medium for solid materials, Rutherford scattering and straggling, or charge exchange when one or more electrons are transferred to the projectile.  

Stoping by charge exchange in p + H and p + D reactions  has been studied in Ref. \cite{BP00}. Because of the low bombarding energy, the electrons adjust themselves quickly and one can use the adiabatic approximation. The potential felt by the electron resembles a symmetric Coulomb barrier around the two atomic centers.  As a function of the relative nuclear distance, the electronic energy levels smoothly undergo from hydrogen to helium levels (left side of Figure \ref{phlev}), passing by molecular orbitals at intermediate distances. Expanding the electronic wavefunction as a time admixture of such orbitals, a set of coupled-channel equations is obtained with help of the Hellmann-Feynman relation. As the bombarding energy decreases only the two lower levels,  $1s\sigma$ and $2p\sigma$ play a relevant role. The electron jumps back and forth between the two atoms by tunneling through the symmetric Coulomb barrier. The right hand side of Figure  \ref{phlev} shows the electron exchange (tunneling) probability as a function of the collision impact parameter in atomic units and for two bombarding energies: (a) 10 KeV, and (b) 100 eV. One notices that at higher energies  the  tunneling probability increases as the distance decreases mainly due to the decrease of the barrier height and the resonant tunneling effect. When the energy is very low, as seen in the lower panel, the electron jumps frenetically back and forth between the two atoms due to resonant tunneling. At very low energies  the exchange probability as a function of impact parameter is given in terms of the  simple formula \cite{BP00}
\begin{equation}
P_{exch} = {1\over 2} +{1\over 2} \cos\left\{ {1\over \hbar}\int_{-\infty}^\infty \left[ E_{2p}(t)-E_{1s}(t)\right]dt \right\}  . \label{pexch}
\end{equation}
The dashed curve in  Figure \ref{phlev} proves that the prediction based on this equation holds very well, specially at very low energies. It is expected to be exact at extremely low energies. The probability minima occur for impact parameters satisfying the relation $\int_{-\infty}^\infty \left[ E_{2p}(t)-E_{1s}(t)\right]dt = 2\pi \hbar (n+1/2)$, a relation more than familiar to all of us. It is  the interference between the $1s\sigma$ and the $2p\sigma$ states that induces oscillations in the exchange probability.

\begin{figure*}
\centering
  \includegraphics[width=0.39\textwidth]{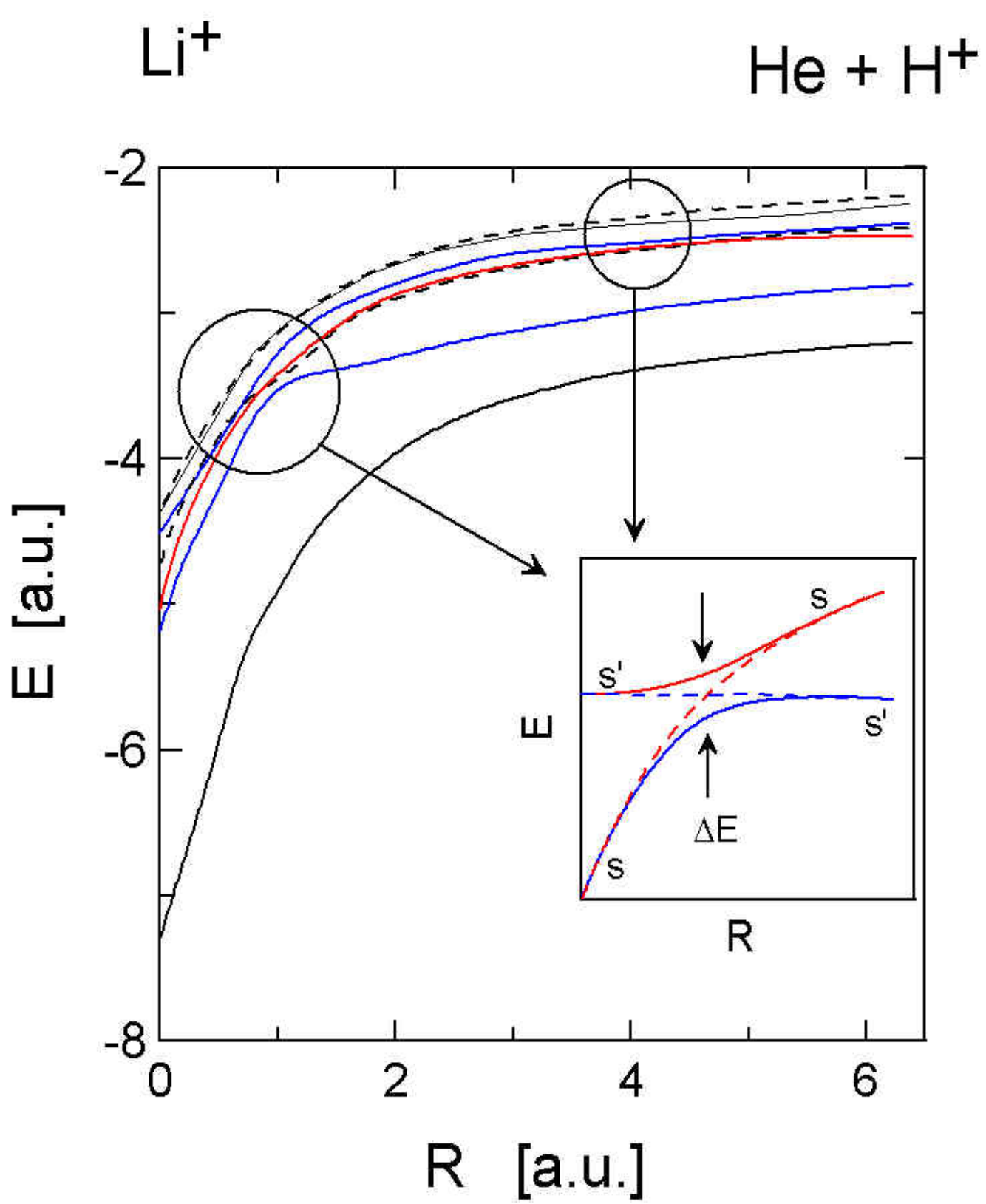}
  \ \ \ \  \ \ \ \ \  \ \ \ \ \ \ \ \includegraphics[width=0.31\textwidth]{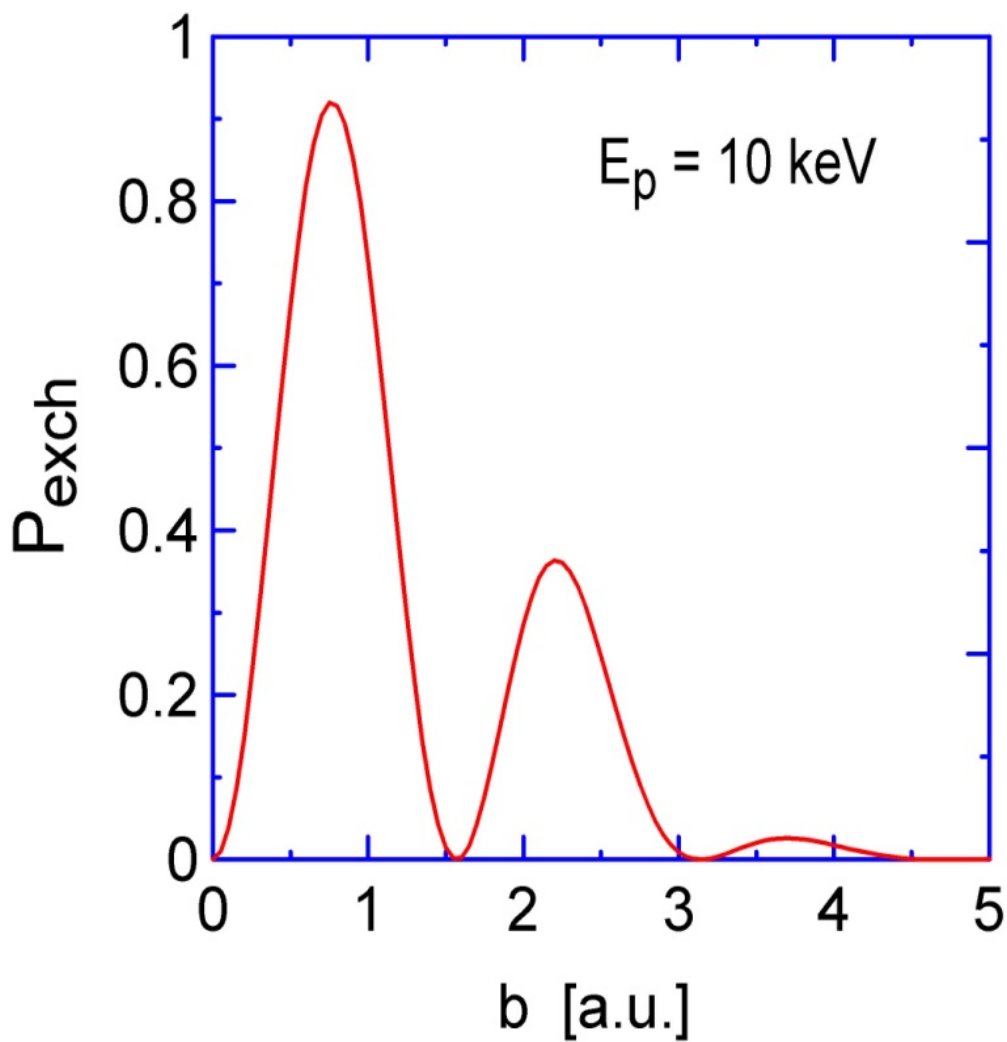}
\caption{{\it Left:} Adiabatic energies (1 a.u. of energy = 27.2 eV, 1 a.u. of length = 0.53 \AA) for the electronic orbitals of the (H-He)$^+$ system as a function of the internuclear separation. As the atoms approach each other slowly, curves of same symmetry repel each other. A transition between states s and s' can occur in a slow collision. In a fast collision a diabatic transition, with the states crossing each other, will occur. This is shown in the inset. {\it Right:} Probability of charge exchange in the collision p + $^4$He showing the damped resonant behavior as a function of the impact parameter and for proton energy $E_p = 10$ keV.}
\label{phelev}       % Give a unique label
\end{figure*}

The case discussed above involves tunneling of only one electron. A more interesting and rather complex situation arises when two electrons are present, as in the case of stopping of protons on a helium gas, as discussed below.

\paragraph{Damping of resonant tunneling - }The case of p + He is more interesting, because of the two active electrons. One can expand the electron wave functions in terms of the two-center basis and obtain a set of time-dependent Hartree-Fock equations \cite{Bert04}. In contrast to the p + H or p + D case the much larger number of levels make a difference. Due to level repulsion, the levels tend to bunch as seen in the left panel of Figure \ref{phelev}.  The transition between the lowest levels shows an oscillating behavior due to resonant tunneling between one electron in the ground state of hydrogen and the first excited state of helium. But the oscillations are  damped. This is due to the interference between the low-lying states and a bunch of states of average energy $\left< E \right>$ and width $2\Gamma$, as shown on the left side of Figure \ref{phelev}. The damping can understood with the Landau-Zener theory for level crossing. At the crossing there is a probability  of an adiabatic transition where $(1-P_{exch})$ given by the Landau-Zener formula
\begin{equation}
P_{exch} = \exp\left[  {2\pi H_{ss'}^2 \over \hbar v d(E_s-E_{s'})/dR} \right] , \label{pexch2}
\end{equation}
where $v$ is the collision velocity, $H_{ss'}$ is the off-diagonal matrix element connecting states $s$ and $s'$ and the term in the denominator is the derivative of the energy difference with respect to the separation distance between the two nuclei.
The interference with the neighboring states introduces a damping in the charge exchange probability, 
\begin{equation}
P_{exch} = \cos^2\left( {\left< E \right> b \over 2\hbar v}\right) \exp\left(- {2 \pi \Gamma^2 b \over \hbar v \left< E \right>}  \right) ,\label{pexch3}
\end{equation}
where $\left< E\right> \approx 1$ a.u. is the average separation energy between the $0\Sigma$ level and the bunch of higher-energy levels shown in Figure \ref{phelev}. The exponential damping factor agrees with the numerical calculations if one uses $\Gamma = 5$ eV. The exchange probability for the electron drops sharply to zero when the energy of the proton is smaller than 18.7 eV. This is easy to understand because this is the energy necessary for the transition $1s^2(^1S_0) \rightarrow 1s2s(^3S)$ in He. Only when this transition occurs, resonant tunneling will be possible between the ground state in hydrogen and the excited state in helium. 

Finally, we consider what happens in tunneling of composite objects through multiple barriers. As an illustrative example, we describe below the case of diffusion and dissociation of molecules in optical lattices. Despite the theoretical simplicity of the system under consideration, the complexity is much larger than in the previous examples and some open questions remain. 

\paragraph{Diffusion in optical lattices - }Using standing wave patterns of reflected laser beams optical lattices can be built to study the diffusion of ultracold atoms and molecules  \cite{LVL11,DZ08,GF08,Sac08}. Fundamental aspects of diffusion of composite objects, such as molecules, subject to transforming transitions from bound to continuum states can be studied in these optical lattices \cite{BBT12}. In Ref. \cite{BBT12} the confining potential described by the optical lattice potential has been assumed to be  a sine function with a periodicity D, equal to half the wavelength of the interfering laser. The Hamiltonian for the molecule is given by $H =T_1 +T_2 +V_1 + V_2 + v$, where $T_i$ is the kinetic energy of atom $i$, $V_i$ is the periodic potential of the lattice and $v$ is the interaction potential between the atoms for a diatomic molecule. A loosely bound, Rydberg-like, system of molecules diffusing through the lattice molecule was studied in Ref. \cite{BBT12}.

\begin{figure*}
\centering
  \includegraphics[width=0.35\textwidth]{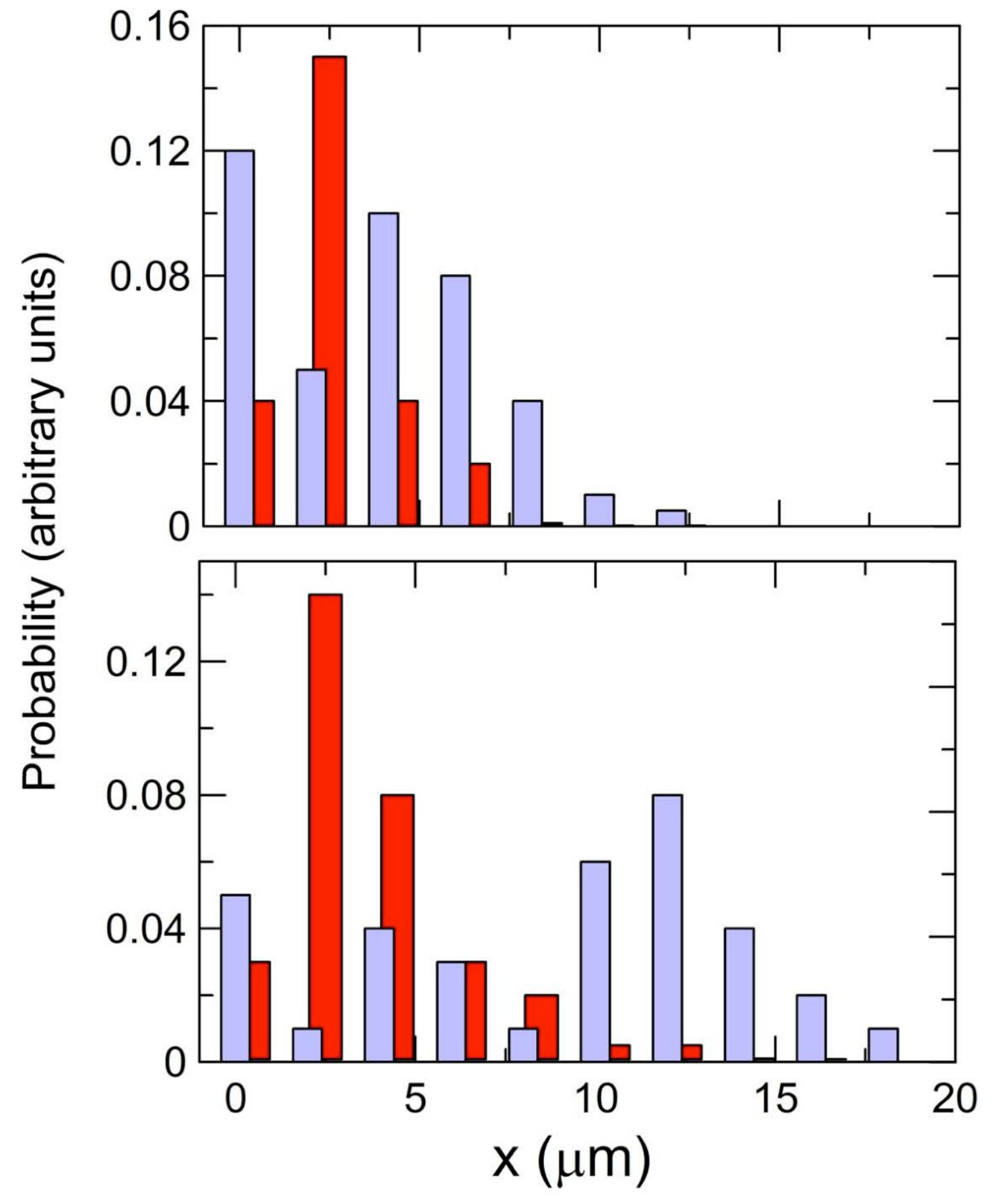}
  \ \ \ \  \ \ \ \ \  \ \ \ \ \ \ \ \includegraphics[width=0.51\textwidth]{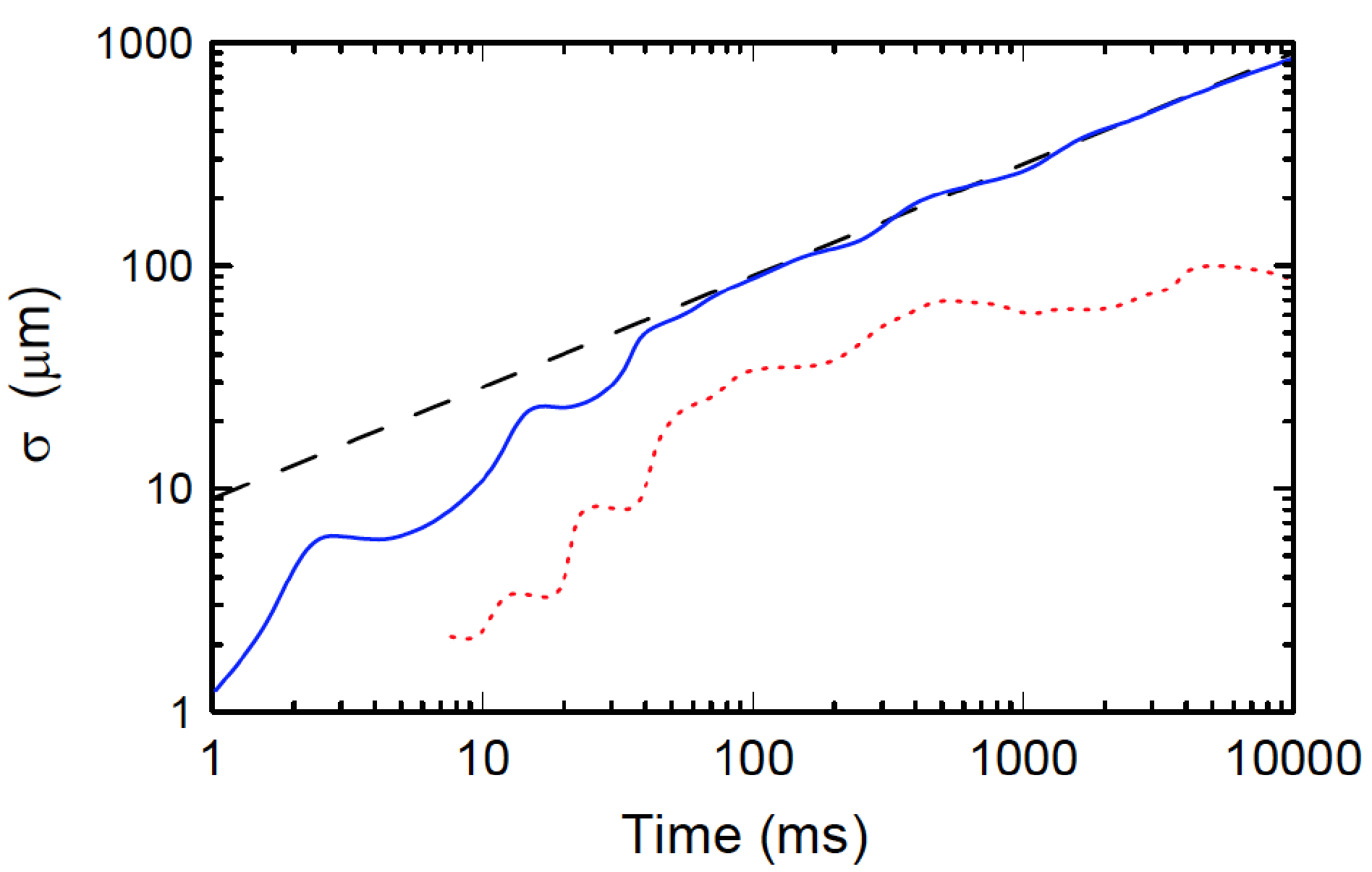}
\caption{{\it Left:} Lattice diffusion of molecules. The hatched histograms are the relative probabilities of finding a molecule in its ground state at a given position along the lattice. The solid histograms give the relative probabilities of finding individual atoms after the dissociation. The molecules are initially confined in the site at the origin. {\it Right:} Time dependence of the spreading width of bound molecules, $\sigma_M (t )$, shown by the solid line, and of dissociated atoms, $\sigma_A (t )$, shown by the dotted line. The dashed curve is a fit of the asymptotic time dependence with the analytical formula, Eq. (\ref{sigdif}).}
\label{diffuse}       % Give a unique label
\end{figure*}

The molecules tend to dissociate as they tunnel through the barrier when their binding energy decreases.  At very small  binding energies, the trend is reversed and the molecule tends to remain intact. As shown in Figure \ref{diffuse}, loosely-bound molecules are resilient to breakup: even at the smallest binding value, the breakup probability is not one.  This result is not unusual, as other tunneling systems exhibit similar behavior. This has been proven for the familiar example of tunneling of Cooper pairs \cite{ZF05}. The pair does not usually dissociate as it tunnels through a barrier. However composite particles may dissociate as the particles tunnel through multiple barriers, as in the case of a lattice potential. It takes time for the molecules to dissociate and additional time for the atoms to diffuse after the dissociation. Therefore, the molecules initially confined within one pocket of the lattice will tunnel and diffuse away from the initial position at a speed higher than that of the atoms that are created in the dissociation of the molecules.

In Ref. \cite{BBT12} a diffusion equation for molecules in an optical lattice  in terms of the diffusion velocity of the wave packet was given by $\sigma(t)=\sqrt{\left< r^2(t) \right>}$, where $\left< r^2(t)\right>$ is the expectation value of the square of the position of the molecule calculated from the solution of the Schr\"odinger equation. The numerical results show that after a transient time, and for a strongly bound molecule of mass $m$, the approximate result holds,
\begin{equation}
\sigma(t)= C \left( \hbar^2 t \over mb \sigma_0^2 \right)^{1/2}, \label{sigdif}
\end{equation}
where $C$ and $b$ are constants depending on the parameters defining the lattice and $\sigma_0=\sigma(t=0)$. $b$ has the meaning of a ``viscosity" generated by the wiggling potential dependence of the lattice. The linear behavior in Eq. (\ref{sigdif}) is shown by the dashed line in Figure \ref{diffuse} (right). Also shown are the numerical results for the same quantity (solid line), and the diffusion velocity of individual atoms in the lattice (dotted line).

A diffusion coefficient can be defined by $D= (\partial \sigma^2/dt)/2$. Using Eq. (\ref{sigdif}) one gets
\begin{equation}
D= {C\over 2} \left( \hbar^2  \over mb \sigma^2_0 \right). \label{sigdif2}
\end{equation}

This result deviates from the classical Einstein diffusion constant, $D = kT/b$. The effective ``temperature" for this system is the kinetic energy of the initial state, $T \simeq \hbar^2/m\sigma^2(t=0)$. The dependence of the temperature, or diffusion coefficient, on the initial kinetic energy is a result which needs to be verified experimentally. The molecules are assumed not to interact. Therefore these results might not hold when molecules interact in a condensate.
The time dependence of the spreading width of the dissociated atoms within the lattice is not well fitted by any power-law dependence on time, as shown in Figure \ref{diffuse}.  The reason for this behavior is not yet understood.

\section{Conclusions}

In this short review I have presented an overview of tunneling of composite objects. I started by giving an example of how one would tackle the problem computationally, followed by an application of the method to the tunneling through a rectangular barrier. It is evident that coupling to open and closed channels lead to a clear resonant behavior visible in the tunneling probabilities.

I went on to discuss a few applications of tunneling of composite objects to real situations, such as molecules passing through thin membranes, cluster-like nuclei tunneling during a nuclear fusion process, and the emission of radiation during tunneling as, e.g., observed during alpha-decay.   Other examples were given in atomic physics where tunneling can enhance or suppress the stopping of particles at very low energies. This important process leads to corrections for the energy loss of ions in reactions of astrophysical interest. Finally, the more difficult theoretical treatment of tunneling of composite objects through multiple barriers has been illustrated with the example of molecules diffusing in optical lattices.

Although a (large) number of theoretical works have studied tunneling phenomena in various situations, quantum tunneling of a composite particle in which the particle itself has an internal structure, has yet to be fully clarified.

\end{document}